\documentclass[%
 preprint,%
 amssymb, amsmath,%
 aip, jap,%
]{revtex4-1}

\usepackage{graphics}
\usepackage{bm}%
\usepackage[colorlinks=true,linkcolor=blue]{hyperref}%
\expandafter\ifx\csname package@font\endcsname\relax\else
 \expandafter\expandafter
 \expandafter\usepackage
 \expandafter\expandafter
 \expandafter{\csname package@font\endcsname}%
\fi
\begin{document}
	
	
	\title{Thermal transport in nanoporous holey silicon membranes investigated with optically-induced transient thermal gratings}
	
	\author{Ryan A. Duncan}
	\email[]{raduncan@mit.edu}
	\affiliation{Department of Chemistry, Massachusetts Institute of Technology, 77 Massachusetts Ave., Cambridge, MA, 02139, United States}
	
	\author{Giuseppe Romano}
	\affiliation{Department of Mechanical Engineering, Massachusetts Institute of Technology, 77 Massachusetts Ave., Cambridge, MA, 02139, United States}
	
	\author{Marianna Sledzinska}
	\affiliation{Catalan Institute of Nanoscience and Nanotechnology (ICN2), CSIC and BIST, Campus UAB, Bellaterra, 08193, Barcelona, Spain}
	
	\author{Alexei A. Maznev}
	\affiliation{Department of Chemistry, Massachusetts Institute of Technology, 77 Massachusetts Ave., Cambridge, MA, 02139, United States}
	
	\author{Jean-Philippe M. Peraud}
	\affiliation{Computational Research Division, Lawrence Berkeley National Laboratory, 1 Cyclotron Road, Berkeley, CA, 94720, United States}
	
	\author{Olle Hellman}
	\affiliation{Department of Applied Physics and Materials Science, California Institute of Technology, Pasadena, CA, 91125, United States}
	
	\author{Clivia M. Sotomayor Torres}
	\affiliation{Catalan Institute of Nanoscience and Nanotechnology (ICN2), CSIC and BIST, Campus UAB, Bellaterra, 08193, Barcelona, Spain}
	\affiliation{ICREA, PG. Llu\'{i}s Companys 23, 08010, Barcelona, Spain}
	
	\author{Keith A. Nelson}
	\affiliation{Department of Chemistry, Massachusetts Institute of Technology, 77 Massachusetts Ave., Cambridge, MA, 02139, United States}

	\date{\today}
	
	\begin{abstract}
		In this study, we use the transient thermal grating optical technique\textemdash a non-contact, laser-based thermal metrology technique with intrinsically high accuracy\textemdash to investigate room-temperature phonon-mediated thermal transport in two nanoporous holey silicon membranes with limiting dimensions of 100 nm and 250 nm respectively. We compare the experimental results to \textit{ab initio} calculations of phonon-mediated thermal transport according to the phonon Boltzmann transport equation (BTE) using two different computational techniques. We find that the calculations conducted within the Casimir framework, i.e. based on the BTE with the bulk phonon dispersion and diffuse scattering from surfaces, are in quantitative agreement with the experimental data, and thus conclude that this framework is adequate for describing phonon-mediated thermal transport through holey silicon membranes with feature sizes on the order of 100 nm.

		
	\end{abstract}
	
	\maketitle

	\section{Introduction}
		Nanoscale thermal transport has become a topic of much recent interest due to the novel transport phenomena that emerge at the micro- and nanoscale \cite{Cahil2003, Cahil2014} and their relevance to technological fields such as microelectronics and thermoelectrics \cite{Shi2012, Vineis2010}. In semiconductor systems with feature sizes comparable to the phonon mean free path (MFP), size effects can lead to strong reductions in thermal conductivity\textemdash making thermal management in microelectronic devices a significant engineering challenge \cite{Pop2006}. In the field of thermoelectrics, nanostructuring has emerged as a key strategy for enhancing the thermoelectric figure of merit $ZT$ by reducing the thermal conductivity without significantly affecting the electronic properties of the material \cite{Vineis2010, Minnich2009}. Traditionally overlooked for thermoelectric applications, silicon has generated recent interest as a material for thermoelectric devices due to the strongly reduced thermal conductivity achievable through nanostructuring \cite{Gadea2018}. Experimental results on silicon nanowires have shown thermal conductivity values two orders of magnitude lower than the bulk value and $ZT$ values approaching unity \cite{Boukai2008, Hochbaum2008, Li2003}. Two-dimensional ``holey silicon'' nanostructures\textemdash suspended silicon membranes with a periodic array of nanopores\textemdash have exhibited thermal conductivity reductions comparable to nanowires \cite{Hopkins2011, Tang2010, Yu2010, Lim2016, Nomura2015, Lee2015} while retaining superior relative mechanical strength. Such nanostructures hold great promise for thermoelectric applications due to the wide variety of well-established and scalable fabrication and manufacturing techniques available for silicon.
		
		Thermal transport at the nanoscale differs significantly from macroscopic, diffusive thermal transport. In structures with feature sizes comparable to the MFP of heat-carrying phonons, thermal transport no longer obeys the  heat diffusion equation \cite{Cahil2003}. One of the earliest attempts to account for non-Fourier phonon-mediated thermal transport in nanostructures was by Casimir \cite{Casimir1938}, whose model featured particle-like phonon transport with diffuse scattering at boundaries. Although Casimir's original model was concerned with thermal transport in rods, the broader formalism of semiclassical particle-like phonon transport with diffuse boundary scattering is expected to be valid for any nanostructure for which $\lambda_{th} \ll \ell$ and $\lambda_{th}/2\pi \lesssim R$ , where $R$ is the surface roughness, $\lambda_{th}$ is the representative wavelength of heat-carrying phonons, and $\ell$  is the limiting dimension of the nanostructure. Heat-carrying phonons at room temperature in silicon have single-digit nanometer wavelengths \cite{Ravichandran2014}, which is on the order of lithographically-realistic surface roughnesses \cite{Saeki2008, Guo2016}. Thus, silicon nanostructures with feature sizes $\ell > 10$ nm should be well described by the Casimir formulation of thermal transport\textemdash that is, particle-like phonon transport according to the phonon Boltzmann transport equation (BTE) with diffuse scattering from surfaces. Studies comparing experimental results with \textit{ab initio} theory based on the BTE have shown that the Casimir formulation is indeed valid for nanoscale silicon membranes \cite{Cuffe2015} and silicon nanobeams \cite{Park2017}. However, there have been highly conflicting reports regarding the validity of the Casimir formulation for thermal transport in nanoporous holey silicon membranes \cite{Ma2014}. Several studies have reported room-temperature effective thermal conductivities reduced by up to an order of magnitude relative to Casimir formulation predictions for such structures \cite{Yu2010, Tang2010, Marconnet2012}, while others have found good agreement between the Casimir formulation and experiment \cite{Jain2013, Verdier2017, Parrish2017}. In some cases, measurements showing quantitative deviations from the Casimir formulation predictions for holey silicon nanostructures have been invoked as evidence of ``coherent'' thermal transport effects at room temperature \cite{Yu2010, Hopkins2011, Alaie2015}. This notion, however, has been challenged by recent experimental and theoretical works \cite{Wagner2016, Maire2017, Lee2016}, in which no effect of nanopore lattice disorder on the room-temperature thermal transport was found. It should be noted that reports of ``below Casimir'' thermal conductivity rely on the measurements of the absolute values of thermal conductivity, which are challenging even for bulk samples \cite{Kremer2004}.  If far-reaching conclusions are to be drawn from the absolute value of thermal conductivity, then a technique with intrinsically high absolute accuracy is desirable.
		
		Transient thermal gratings (TTG) is a non-contact optical technique that measures the time evolution of an impulsively generated sinusoidal temperature profile \cite{Johnson2012, Vega-Flick2016}. The experimental observable is the amplitude of this sinusoidal temperature profile, which decays as heat spreads from the peaks to the nulls of the grating. For a one-dimensional TTG, the amplitude of the thermal profile and therefore the intensity of the heterodyned TTG signal is given by
		
		\begin{equation}
			I(t) \propto e^{-t/\tau}
			\label{eq:TTG_sig}
		\end{equation}
		
		where $\tau \equiv 1/\alpha q^2$, $\alpha$ is the thermal diffusivity, $q \equiv 2\pi/L$ is the transient grating wavevector, and $L$ is the transient grating period. The only parameter other than $\alpha$ that affects the decay rate is $L$, which can be measured with high accuracy. Thus the thermal diffusivity can be determined to high accuracy from the decay rate of the TTG signal.  Furthermore, TTG's non-contact nature reduces additional sources of error due to the absence of any interfaces with metrological structures.
		
		In this paper, two 250 nm-thick holey silicon membrane nanostructures are investigated with the TTG technique. The experimental results from TTG measurements are compared to the results of two \textit{ab initio} numerical Boltzmann transport techniques: the OpenBTE computational framework developed by Romano et al. \cite{Romano2015} and the energy-based deviational Monte Carlo BTE simulation technique developed by Peraud and Hadjiconstantinou \cite{Peraud2011, Peraud2012}. Quantitative agreement between numerical calculations and experiment is found for both the unpatterned silicon membrane and holey silicon structures, confirming the validity of the Casimir formulation for room temperature heat transport in  silicon nanostructures with feature sizes on the order of 100 nm.
	
	\section{Experimental}
		\subsection{Sample Fabrication}
\begin{figure}
				\includegraphics{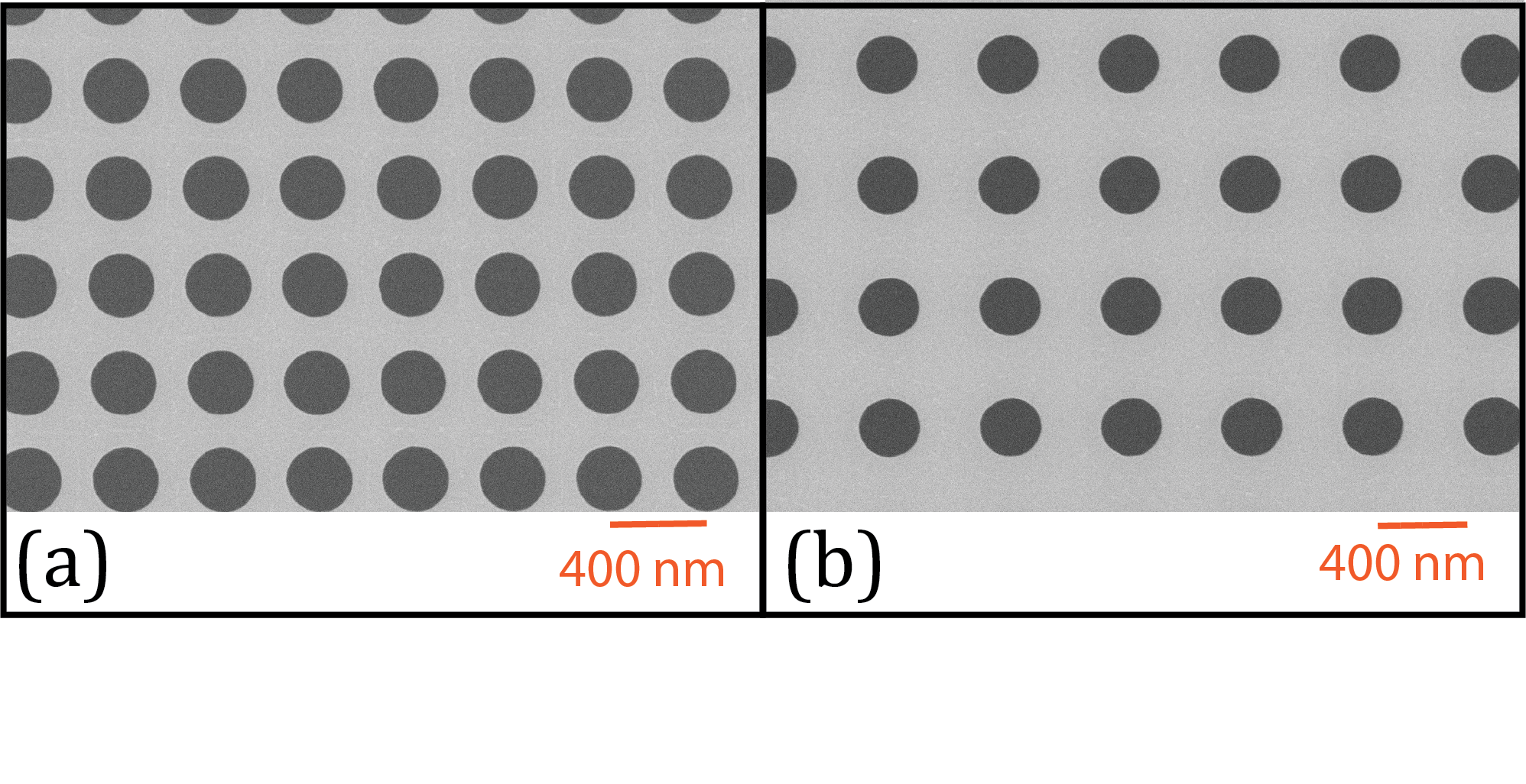}
				\caption{Scanning electron micrographs of the patterned holey silicon membranes\textemdash (a) region A (400 nm pitch, 280 nm nanopore diameter), and (b) region B (500 nm pitch, 250 nm nanopore diameter).}
				\label{fig:SEM}
			\end{figure}

			The holey silicon structures were fabricated using electron beam lithography (EBL) and reactive ion etching (RIE) of a 250 nm-thick freestanding silicon membrane $3.2 \times 3.2$ $\mu$m window area, obtained from Norcada Inc.) \cite{Sledzinska2016}. Each of the two structures was a 100 $\mu$m-diameter region of the freestanding membrane patterned with a square lattice of nanopores. SEM micrographs of the regions are shown in Fig. \ref{fig:SEM}. ``Region A'' had a pitch size (nanopore periodicity) of 400 nm and a nanopore diameter of 280 nm, and ``region B'' had a pitch size of 500 nm and a nanopore diameter of 250 nm.
			
			\subsection{Transient thermal grating (TTG) measurements}
			In TTG, two ``pump'' laser pulses are crossed at the sample, where optical interference and subsequent absorption lead to the establishment of a transient sinusoidal temperature profile with spatial period $L = \lambda / 2 \sin{(\theta/2)}$, where $\lambda$ is the pump wavelength and $\theta$ is the crossing angle for the two pump beams. Through the temperature dependence of the material's complex index of refraction $\tilde{n} \equiv n + ik$\textemdash where $n$ is the real index of refraction and $k$ is the absorption coefficient\textemdash this sinusoidal temperature profile is accompanied by a spatially sinusoidal modulation in $\tilde{n}$ as well. A quasi-continuous ``probe'' beam then impinges on the sample, diffracting from this transient optical grating. As the amplitude of the temperature grating diminishes due to heat transport from the peaks to the troughs, the amplitude of the grating in $\tilde{n}$\textemdash and thus the amplitude of the diffracted signal\textemdash diminishes accordingly. In this way, the time dependence of the diffracted signal can be directly related to the thermal diffusivity according to Eq. \ref{eq:TTG_sig}. TTG measures the thermal transport dynamics over a length scale set by the period of the transient grating, which can be tuned by changing the crossing angle of the pump beams. Further details regarding this technique can be found in Ref. \cite{Vega-Flick2016}.
			
			The pump beams were derived from a 515 nm source with a 60 ps pulse duration and 1 kHz repetition rate, and the probe beam was derived from a continuous-wave 532 nm source. A ``reference'' beam was derived from the same source as the probe beam, and the relative phase between the two was controlled by tilting a highly parallel optical flat through which the probe beam passes to achieve heterodyne detection \cite{Maznev1998}. At the sample the probe beam diffracts from the transient grating and becomes superposed with the transmitted reference beam, and the combined heterodyned signal is collected by a fast photodiode detector and recorded on an oscilloscope. The $1/e^2$-intensity radius of the pump and probe beams were 100 $\mu$m and 40 $\mu$m, respectively. While the pump spot size is commensurate with the patterned regions, the probe spot size is much smaller. Thus, although our pump may be exciting a grating pattern that extends somewhat outside of the patterned region, our experiment is only sensitive to the transport dynamics within the region bounded by the much smaller probe spot. The pump pulse energies of the measurements ranged from 170 - 340 nJ, and the instantaneous power of the probe beam at the sample ranged from 0.8 - 1.6 mW. The probe beam was shuttered by an electro-optic modulator with a duty cycle of 5\% to prevent steady-state heating of the sample.
			
\begin{figure}
				\includegraphics{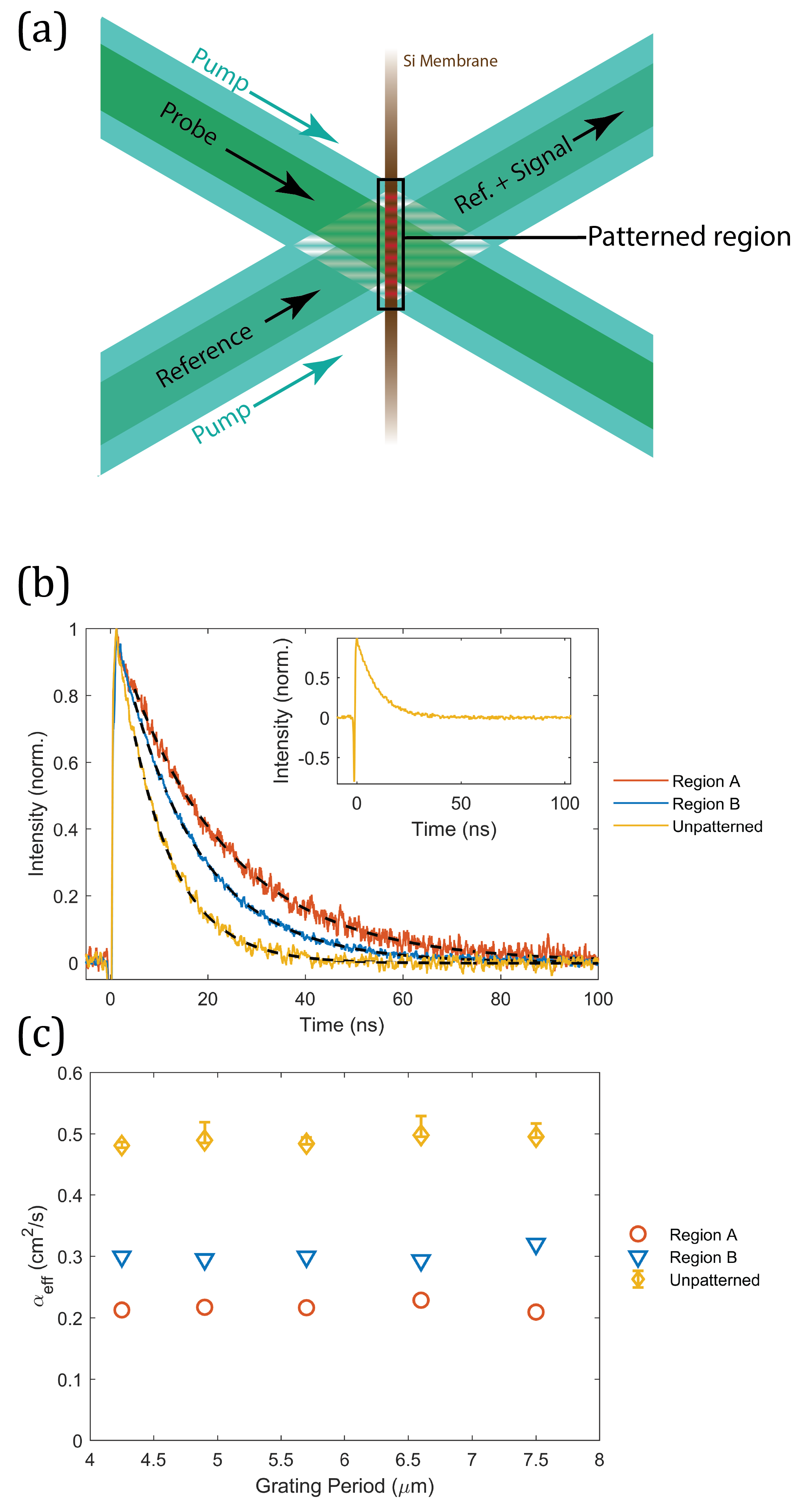}
				\caption{(a) Schematic of the TTG measurement in the transmission geometry. (b) Time-domain TTG traces at 4.25 $\mu$m transient grating period for the two holey silicon membranes and the unpatterned silicon membrane, as well as exponential fits to the data. Inset: full transient grating response for the unpatterned membrane, including fast early-time electronic signal. (c) Effective thermal diffusivity values obtained from the time-domain data according to Eq. \ref{eq:TTG_sig}. For the patterned regions the error is smaller than the sizes of the symbols.}
				\label{fig:TTG}
			\end{figure}

			The TTG measurements in this work were performed in a transmission geometry since the thickness of the membrane is smaller than the optical penetration depth of silicon for the wavelengths involved in the measurements. A schematic of the experimental geometry is shown in Fig. \ref{fig:TTG}(a), and the raw TTG data obtained from the two holey regions and the unpatterned silicon membrane at a grating period of 4.25 $\mu$m are shown in Fig. \ref{fig:TTG}(b). Measurements were performed under medium vacuum at a pressure of 1 mbar. The maximum amplitude of the temperature grating was determined to have an upper bound of 35 K. Upper bounds on the average heating of the sample due to the pump and probe beams were determined to be 20 K each.
			
			The TTG signal for a one-dimensional thermal grating exhibiting diffusive thermal transport is given by Eq. \ref{eq:TTG_sig}. $\tau$\textemdash or equivalently $\alpha$\textemdash is the only free parameter required to model the normalized TTG signal. In addition, the heterodyne detection scheme further yields a gain in signal-to-noise by a factor of the reference field amplitude, which can be increased arbitrarily up to the saturation threshold of the photodetector. The low-dimensionality of the dynamical parameter space, the signal gain provided by the reference field, and the fact that neither precise knowledge of the magnitude of the temperature variation nor of the magnitude of the heat flux is required in the analysis of the data allow for the determination of the thermal diffusivity with high absolute accuracy. Further discussion regarding the accuracy of transmission-geometry TTG experiments on nanomembranes can be found in Ref. \cite{Vega-Flick2016}. The traces were truncated such that fitting began 5 ns after pump incidence to ensure that the fitted region corresponds only to thermal transport signal without any potential contribution from the fast electronic response shown in the inset of Fig. \ref{fig:TTG}(b). The acquired fits are plotted alongside the raw TTG data in Fig. \ref{fig:TTG}(b).  Fig. \ref{fig:TTG}(c) shows the measured thermal diffusivity values obtained according to Eq. \ref{eq:TTG_sig} as a function of TTG period for each of the three regions measured. Each raw TTG trace consisted of 50,000 individual measurements. The statistical error of the measurement was determined by partitioning the data into subsets of 10,000 measurements, fitting each subset to Eq. \ref{eq:TTG_sig}, and taking the standard error of the mean of the resulting distribution of $\tau$ values. In addition to the statistical error of the measurement, the systematic error due to laser heating effects was also considered. The effects of laser heating were determined by performing each measurement three times\textemdash once at a baseline set of pump and probe powers, and two additional times at which the pump and probe powers respectively were doubled. Linearly extrapolating the measured values of $\tau$ to zero pump and probe laser power allows us to determine the systematic error due to laser heating, which was then added to the appropriate side of the errorbars for each point to account for this systematic heating effect. We note that the upper bounds on laser heating reported above are non-negligible relative to room temperature. However, since the effect of laser heating is experimentally quantified in our error analysis, we can still compare our experimental results with calculations that use room-temperature material properties. Despite the somewhat high upper bounds on laser heating, we nevertheless note that the effect of laser heating on the experimentally-determined values of $\alpha$ was found to be only $\sim10\%$ or less.
			
			For grating periods from 4.25-7.5 $\mu$m we find that the experimental values of thermal diffusivity are independent of $L$ for both the unpatterned membrane and the holey membranes, consistent with preliminary TTG results on holey silicon structures \cite{Vega-Flick2016}. The exponential form of the TTG data and the invariance of thermal diffusivity as a function of grating period indicates that the transport kinetics are ``effectively diffusive'' over the TTG experimental length scales, albeit with ``effective'' thermal diffusivity values $\alpha_{eff}$ reduced relative to the bulk because of the non-Fourier size effect due to nanostructuring.
			
			It should be noted that occasionally an additional transient with a characteristic timescale much longer than the acquisition timescale (i.e., approximately a constant offset from the pre-pump baseline) was observed in some of the obtained TTG traces. However, we determined that the presence of this contribution to the signal (which is roughly on the timescale that would correspond to thermal diffusion out of the pump spot) was not associated with any change in the $\alpha_{eff}$ value that was calculated from the time constant of the exponentially decaying contribution to the signal observed on the 10s-100s ns timescale (which we took to be the true TTG signal) that remained after subtracting out this approximately constant offset. This issue is more thoroughly addressed in the Supplementary Material.
			
\begin{figure}
				\includegraphics{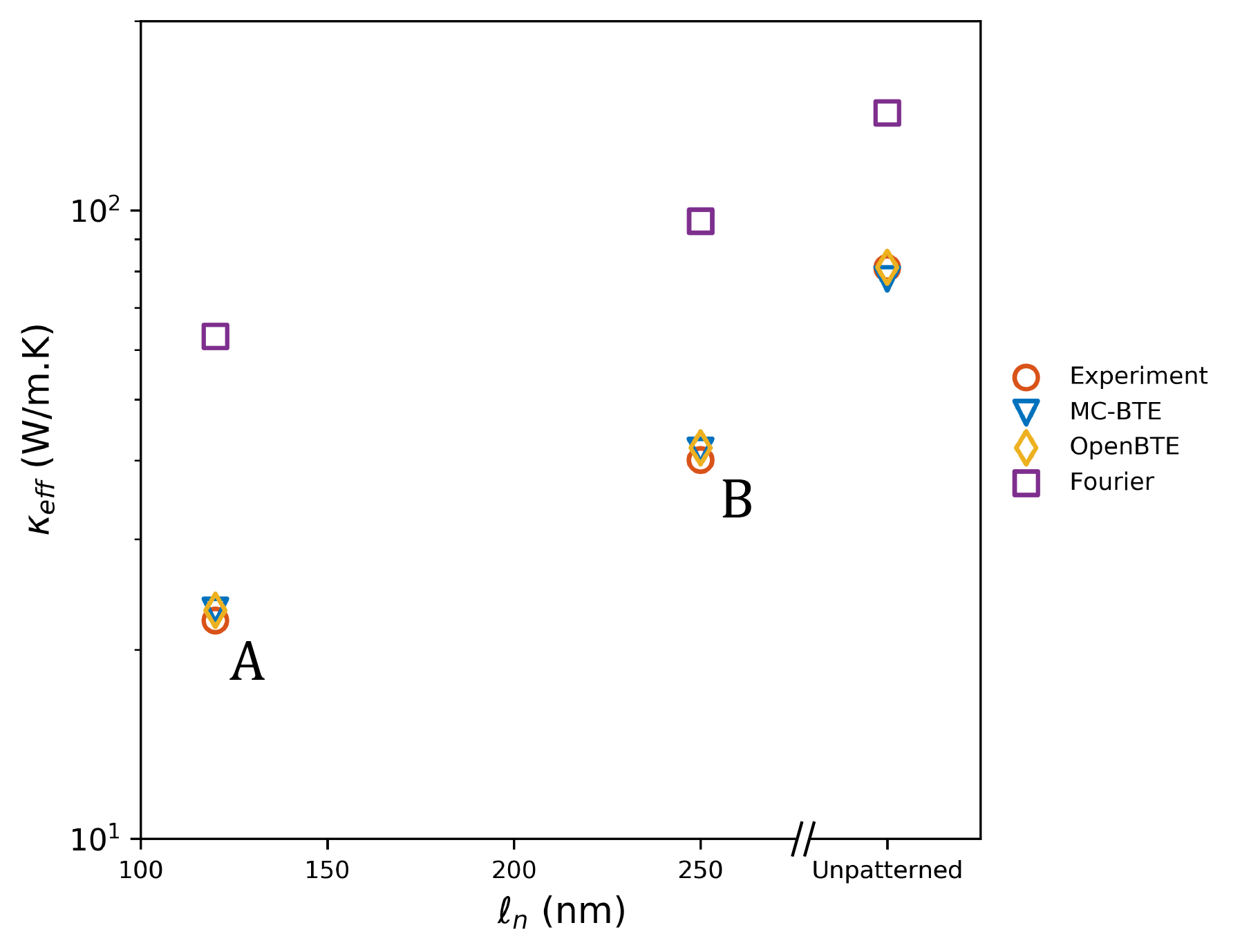}
				\caption{Effective thermal conductivity values experimentally measured and numerically computed using the MC-BTE and OpenBTE methods for the holey silicon regions and the unpatterned membrane. Also plotted are the $\kappa_{eff}$ values obtained by using the Fourier law with the bulk silicon thermal conductivity.  $\ell_{n}$ is the neck width. The error was determined to be smaller than the size of the symbols for both the experimental and (all) computational results.}
				\label{fig:kappa}
			\end{figure}

			Experimental values of the effective thermal conductivity $\kappa_{eff}$ were calculated from the data in Fig. \ref{fig:TTG}(c) according to
			
			\begin{equation}
				\kappa_{eff} = (1 - \phi)c_{Si} \alpha_{eff}
				\label{eq:kappa_eff}
			\end{equation}
			
			where $\phi$ is the void fraction of the holey silicon membrane and $c_{Si}$ is the bulk volumetric specific heat of silicon. The resulting experimental values of $\kappa_{eff}$ are shown in Fig. \ref{fig:kappa}, where the effective thermal conductivity values are plotted against the neck width $\ell_{n}$ (i.e., the difference between the pitch size and the nanopore diameter).
			
	\section{Comparison to first-principles numerical calculations}
		Numerical calculations of the thermal transport through the membranes were performed according to the linearized isotropic phonon Boltzmann transport equation (BTE) under the single-mode relaxation time approximation (RTA), which is given by
		
		\begin{equation}
			\frac{\partial f_{\mathbf{k} p}}{\partial t} + \mathbf{v}_{\mathbf{k} p} \cdot \mathbf{\nabla}f_{\mathbf{k} p} = \frac{f_0  - f_{\mathbf{k}p}}{\tau_{kp}}
			\label{eq:BTE}
		\end{equation}
		
		where $f_{\mathbf{k} p}(\mathbf{r},t)$ is the occupation function for a mode traveling with wavevector $\mathbf{k}$ and polarization $p$, $\mathbf{v}_{\mathbf{k} p}$ is the (isotropic) group velocity (where $k\equiv|\mathbf{k}|$), $f_0 \big(\hbar \omega,T_L (\mathbf{r},t)\big)$ is the Bose-Einstein distribution, $T_L(\mathbf{r},t)$ is the local temperature field defined such that energy is locally conserved, $\hbar \omega$ is the phonon energy, and $\tau_{kp}$ is the (isotropic) single-mode relaxation time.

		The simulation domain is one pore-centered unit cell of the nanopore lattice with the cylindrical axis of the pore chosen to be oriented along $\mathbf{\hat{z}}$. Periodic boundary conditions are applied along both the $x$- and $y$-axes. The phonon group velocities and relaxation times were determined respectively from the harmonic and anharmonic force constants, which were obtained from density functional theory (DFT) calculations using the temperature dependent effective potential (TDEP) method \cite{Hellman2013}. Naturally occurring isotope disorder was taken into account. Details on the DFT calculations can be found in the Supplementary Material.
		
		The OpenBTE computational technique of Romano et al.~\cite{Romano2015} and the energy-based deviational Monte Carlo BTE (MC-BTE) technique of Peraud and Hadjiconstantinou~\cite{Peraud2011, Peraud2012} were both used for \textit{ab initio} calculations of $\kappa_{eff}$ for both the holey and unpatterned membranes.

		For the OpenBTE case, Eq.~\ref{eq:BTE} is transformed into the following form~\cite{Romano2015}:
		
		\begin{equation}
			\begin{split}
			\Lambda \mathbf{\hat{s}}(\Omega) \cdot \mathbf{\nabla}T(\mathbf{r},\Omega,\Lambda) + T(\mathbf{r},\Omega,\Lambda) = T_L (\mathbf{r}), \\
			T_L = \bigg[ \int^{\infty}_{0} \frac{K(\Lambda')}{\Lambda'^2} \textrm{d} \Lambda ' \bigg]^{-1} \int^{\infty}_{0}\frac{K(\Lambda'')}{\Lambda''^2} \langle T(\mathbf{r},\Omega,\Lambda'')  \rangle \textrm{d} \Lambda''
			\end{split}
			\label{eq:OpenBTE}
		\end{equation}
		
		where $\mathbf{\hat{s}}(\Omega)$ is the unit vector for the propagation direction $\Omega$, $T(\mathbf{r},\Omega,\Lambda)$ is the ``effective temperature'' of phonons with MFP $\Lambda$ traveling in direction $\Omega$ (i.e., the sum of their energy densities divided by $c_{Si}$),  $K(\Lambda)$ is the bulk MFP distribution (i.e., the derivative of the thermal conductivity accumulation function with respect to $\Lambda$), and $\langle x(\Omega)\rangle \equiv (1/4\pi) \int_{4\pi} x{\Omega} \mathrm{d}\Omega $ is the angular average over all propagation directions. Eq. \ref{eq:OpenBTE} is derived by imposing steady-state conditions on Eq. \ref{eq:BTE}, as well as assuming that $\delta T(\mathbf{r}) \equiv T_L (\mathbf{r})-T_0$ (where $T_0$ is the reference temperature, which in this study was 300 K) is small such that $f_0 [\hbar \omega,T_L (\mathbf{r})]$ in Eq. \ref{eq:BTE} can be expanded to first order in $\delta T(\mathbf{r})$ and any temperature dependence of material properties can be neglected. The advantage of this approach is that the only input required to solve Eq. \ref{eq:OpenBTE} is the MFP distribution $K(\Lambda)$.

		In OpenBTE, a difference of temperature $\Delta T$ is applied at the two opposing faces of the unit-cell along the $x$-axis, and the first guess for $T_L$ was given by the standard diffusive equation. Diffuse scattering at boundaries was modeled in such a way that phonons of a given value of $\Lambda$ were diffusely emitted (i.e., emitted equally in all directions) from the surface with a total energy equal to the total energy of all incident  particles with the same value of $\Lambda$. To overcome numerical instability due to small-MFP phonons, OpenBTE switches to a modified Fourier's law to compute the diffusive component to heat transport~\cite{Romano2019} for such modes. Totally diffuse-scattering boundary conditions were imposed on all surfaces of the computational domain. Eq.~\ref{eq:OpenBTE} was solved by the finite-volume method while a Delaunay mesh was generated for space discretization \cite{Murthy1998,romano2011multiscale}.
		
		
		
		The deviational energy-based MC-BTE technique \cite{Peraud2011, Peraud2012} was also used to calculate $\kappa_{eff}$ for the nanostructures investigated. This technique achieves low statistical variance compared to other Monte Carlo techniques by only simulating the trajectories of ``deviational'' particles which describe the excess/deficit thermal energy in a given mode relative to equilibrium, and achieves high computational efficiency by performing the calculation in an energy-based BTE formulation that lends itself naturally to energy conservation. The diffuse boundary scattering condition was modeled in the same fashion as in the OpenBTE method described above\textemdash namely, deviational particles with a given MFP were diffusely emitted from the surface with a total energy equal to the total energy of all incident particles with the same MFP. Unlike the OpenBTE technique as described in Ref. \cite{Romano2015}, the MC-BTE solver applies a constant temperature gradient throughout the simulation domain rather than isothermal conditions at the $\mathbf{\hat{x}}$-normal boundaries. To assess the impact of this discrepancy in the applied perturbation we compute $\kappa_{eff}$ with OpenBTE using both approaches on test aligned structures, finding negligible differences in effective thermal conductivity values.
		
		For both computational techniques, the conductance of one unit cell was calculated by dividing the total heat flux through one end of the simulation domain by $\Delta T$. The effective thermal conductivity $\kappa_{eff}$ was then obtained by dividing this conductance value by the rectangular cross-sectional area of the unit cell normal to $\mathbf{\hat{x}}$ and by multiplying by the unit cell length along $\mathbf{\hat{x}}$ \cite{note1}. Our results for both computational techniques are shown in Fig. \ref{fig:kappa} for comparison to the experimental TTG results for both holey silicon structures and the unpatterned membrane.

		In a previous paper \cite{Vega-Flick2016}, preliminary TTG results investigating thermal transport in a similar holey silicon membrane were compared to the values of $\kappa_{eff}$ obtained from \textit{ab initio} MC-BTE simulations. Agreement between experiment and theory was found to within $\sim 20\%$. However the sample in that study was patterned over the entirety of the suspended membrane, and as such comparison with an unpatterned region to ensure the intrinsic quality of the silicon membrane was not possible. There thus remained an ambiguity in the previous study as to whether the discrepancy between theory and experiment was due to deviations from the Casimir formulation, or due simply to material quality effects. Our computational results for $\kappa_{eff}$ are shown alongside our experimental results in Fig. \ref{fig:kappa}, as well as the values of $\kappa_{eff}$ for the structures calculated using the Fourier law with the bulk silicon thermal conductivity value of 143 W/m.K. We see that the size effect associated with the thickness of the unpatterned membrane alone reduces $\kappa_{eff}$ by nearly a factor of two relative to the value obtained from the Fourier law (which is simply the value for bulk silicon in the case of the unpatterned membrane), in good agreement with previous measurements on nanoscale silicon membranes \cite{Cuffe2015}. The quantitative agreement between the experimental and computational results for an unpatterned region ensures the intrinsic sample quality of the membrane, and allows us to associate any deviation between experimental and computational results for the patterned regions solely with the introduction of the nanopore lattice. A further reduction of $\kappa_{eff}$ is observed due to the nanopore superlattice patterning, resulting in a reduction of $\kappa_{eff}$ by a factor of 3 relative to the Fourier law prediction for region A and a near order of magnitude reduction in $\kappa_{eff}$ for region A relative to bulk silicon. The quantitative agreement that we obtain between first-principles BTE computational techniques and non-contact high-accuracy TTG thermal transport measurements indicate that the broader Casimir formulation for lattice-based thermal transport is indeed quantitatively accurate for treating room-temperature thermal transport through periodic holey silicon membranes with feature sizes on the order of 100 nm.

	\section{Conclusions}
		We have used the non-contact optical TTG method to investigate thermal transport in two nanostructured holey silicon membranes. We observe effectively diffusive transport at grating periods larger than 4 $\mu$m and a reduction in effective thermal conductivity by nearly an order of magnitude relative to the bulk value. Two \textit{ab initio} numerical techniques simulating transport according to the semiclassical phonon Boltzmann transport equation yielded excellent agreement with the measurements. Our results indicate that the Casimir framework of semiclassical particle-like phonon-mediated thermal transport with diffuse boundary scattering is adequate for describing thermal transport in holey silicon structures with limiting dimensions of $\sim$100 nm.

	\section*{Acknowledgements}
		We would like to thank Charles Shi and Jonas Rajagopal for assisting with the TTG measurements. RAD and KAN acknowledge support from the NSF EFRI 2-DARE grant EFMA-1542864. AAM and KAN acknowledge support by the Solid State Solar-Thermal Energy Conversion Center (S3TEC), an Energy Frontier Research Center funded by the U.S. Department of Energy, Office of Science, Office of Basic Science, under award DE-SC0001299. The ICN2 is funded by the CERCA program / Generalitat de Catalunya and supported by the Severo Ochoa Centres of Excellence program, funded by the Spanish Research Agency (AEI, grant No. SEV-2017-0706). MS and CMST acknowledge support from the Spanish National Project SIP (PGC2018-101743-B-100) and from AGAUR Grant No. 2017SGR806. OH gratefully acknowledges financial support from the VINN Excellence Center for Functional Nanoscale Materials (FunMat-2) Grant 2016-05156 and the Knut and Alice Wallenberg Foundation through Wallenberg Scholar Grant No. 2018.0194. This work has been partly supportedby La Caxia Foundation MISTI Global Seed Fund program (LCF/PR/MIT18/11830008).
		
	\section*{Data Availability Statement}
		The data that support the findings of this study are available from the corresponding author upon reasonable request.
		
	\bibliography{Holey_Si_bibliography_review}

\begin{thebibliography}{51}
\bibitem [{not()}]{note1}%
  \BibitemOpen
  \href@noop {} {}\bibinfo {note} {Thus $\kappa_{eff}$ is defined as the
  thermal conductivity of a solid membrane that would have the same conductance
  per unit area as the nanoporous membrane. This definition is consistent with
  Eq. \ref{eq:kappa_eff} used to obtained the experimental value of
  $\kappa_{eff}$ from the measured thermal diffusivity.}\BibitemShut {Stop}%
\end{thebibliography}
	
	\renewcommand{\thefigure}{S\arabic{figure}}
	\setcounter{figure}{0}

	\section*{Supplementary Material}
	
		\subsection*{Long-time, approximately constant contribution to the signal during TTG measurements}
\begin{figure}[h!]
			\includegraphics{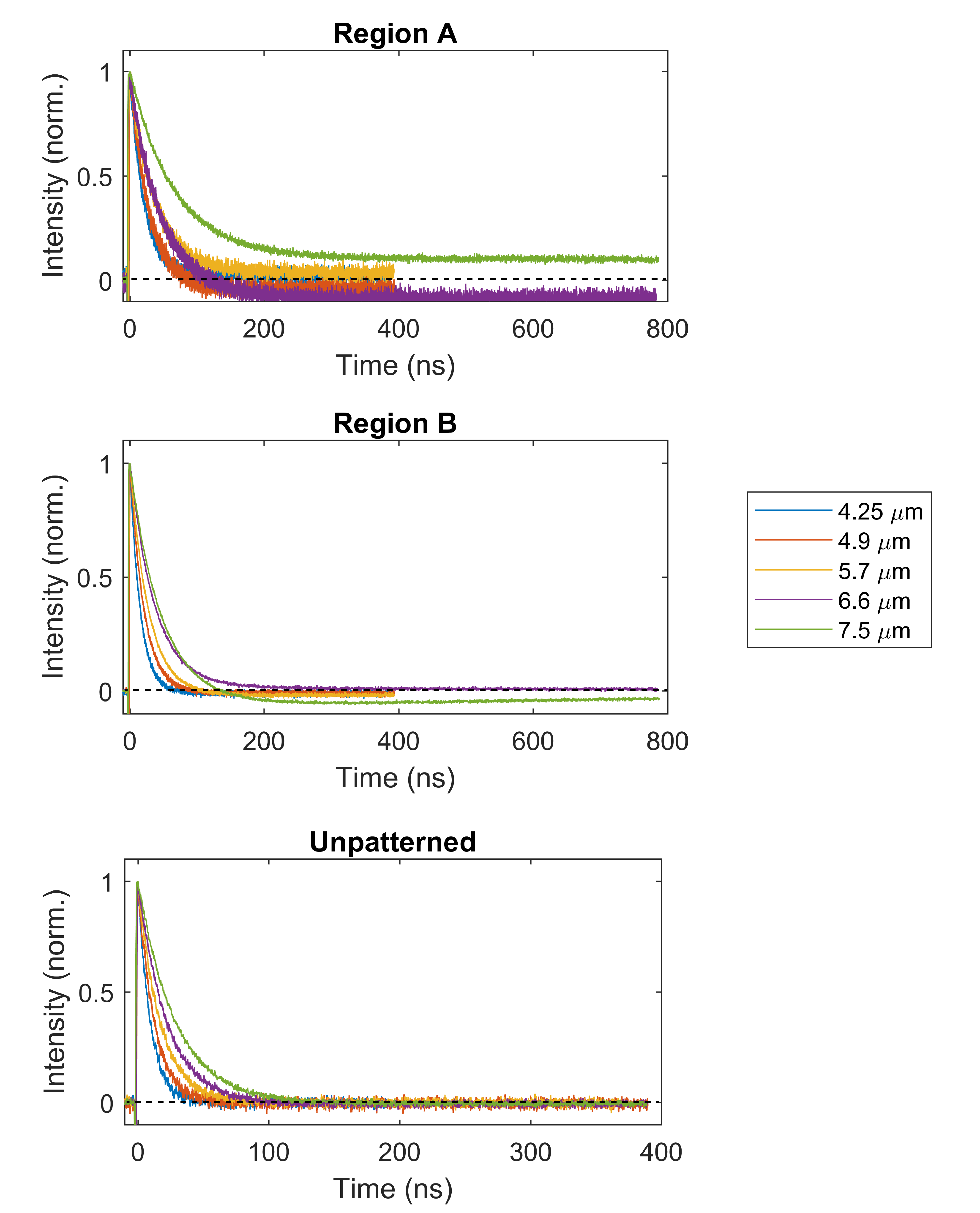}
			\caption{Normalized TTG traces obtained for all regions at every grating period measured, with baselines set to the pre-pump values.}
			\label{fig:baseline_traces}
		\end{figure}

\begin{figure}[h!]
			\includegraphics{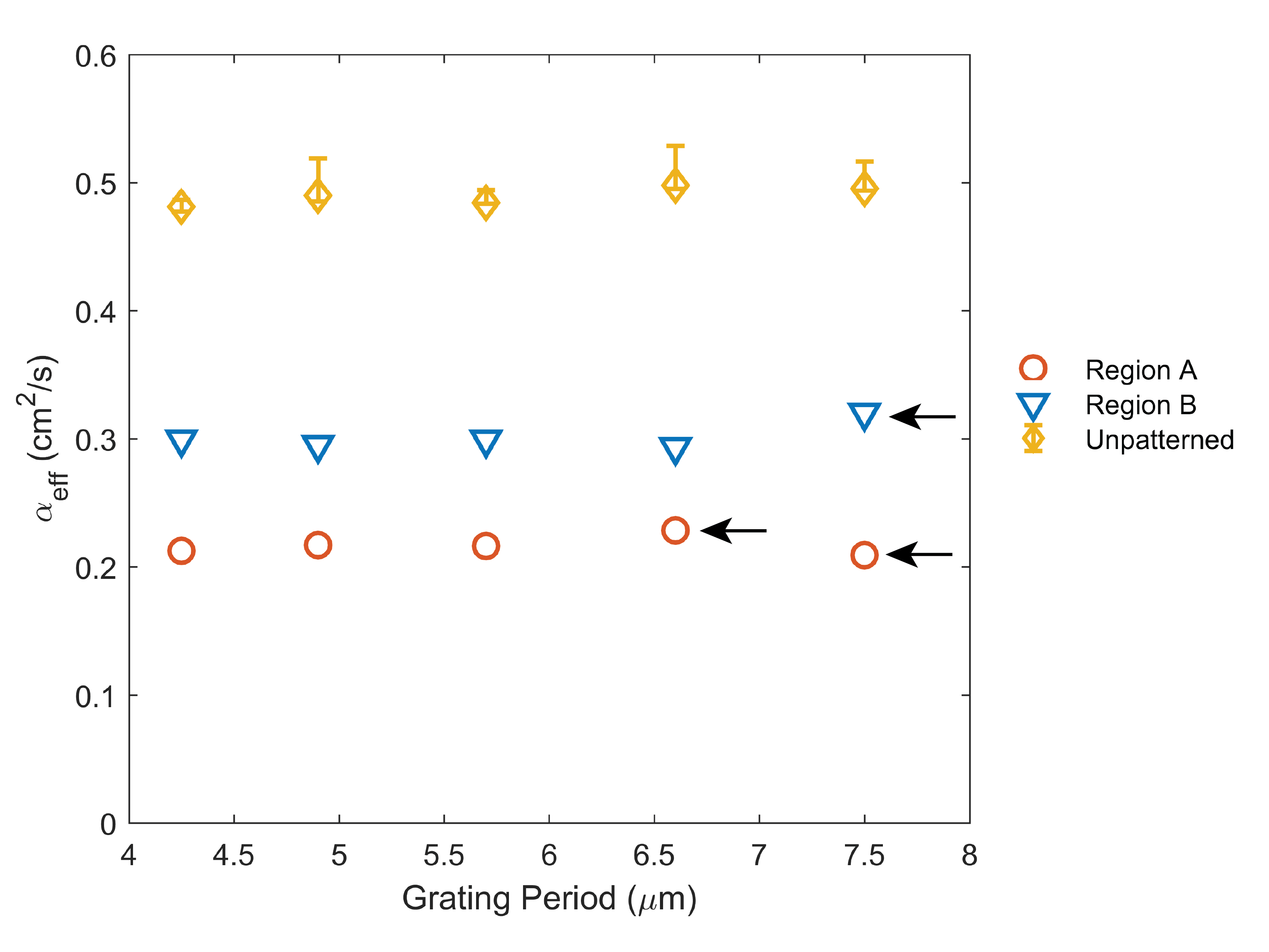}
			\caption{$\alpha_{eff}$ values determined from the data in Fig. S1, where arrows correspond to measurements where the signal at the end of the acquisition time window remains $>$ 5\% maximum amplitude away from the pre-pump baseline values. We see that our determined values of $\alpha_{eff}$ are independent of the presence of this long-time contribution to the signal for all regions.}
			\label{fig:baseline_alpha}
		\end{figure}

		For some of the measurements performed in this work a contribution to the measured signal at much longer timescales than the normal TTG decay was observed. The raw TTG traces for all regions investigated at all grating periods are plotted in Fig. \ref{fig:baseline_traces}, where it can be seen that the signal in some of the measurements of the patterned regions does not decay to the pre-pump baseline by the end of the TTG decay. We see that this contribution is present in the holey silicon regions at the larger grating periods studied\textemdash i.e., grating periods of 6.6 $\mu$m and 7.5 $\mu$m for region A, and 7.5 $\mu$m alone for region B. Interestingly, this long-time signal contribution does not appear in any of the TTG measurements of the unpatterned membrane. The timescales of these very slow transients (10s - 100s of $\mu$s\textemdash much longer than the acquisition time window used to capture the entirety of the ``true'' TTG signal but shorter than the time between pump pulses) are roughly consistent with thermal diffusion of the deposited heat at a diffusivity of $\alpha_{eff}$ out of the 100 $\mu$m-diameter pump spot. However, it is not clear why this contribution would be present at some grating periods while not in others, nor is it clear why such signal would be present in the heterodyned TTG signal at all. This very slow contribution to the signal is well-separated in terms of timescale from the faster decay on 10s-100s ns timescales (which we take to be the ``true'' TTG signal corresponding to thermal transport from the peaks to the nulls of the thermal grating), and we choose to treat it as a constant offset when fitting the faster decay to determine $\alpha_{eff}$. Fig. 2(c) is recreated in Fig. \ref{fig:baseline_alpha}, where the measurements corresponding to TTG traces which did not decay to the baseline over the acquisition time window are indicated with arrows. We see that the presence of this long-time signal does not have any appreciable effect on the $\alpha_{eff}$ values obtained, which indicates that it is a separate and independent contribution to the signal that has no impact on the signal arising from thermal transport from the peaks to the troughs of the transient grating.
		
		\subsection*{Detail on density functional theory calculations}
		
		Parameters for the lattice dynamical calculations were obtained from DFT calculations as implemented in VASP\cite{kresse1993ab,kresse1999ultrasoft,kresse1996efficiency,kresse1996efficient}. The calculations used a 5 x 5 x 5 supercell, a 500 eV plane wave energy cutoff and exchange correlation was treated with the AM05 functional~\cite{mattsson2009implementing,armiento2005functional}. The phonon mean free paths were calculated at $T = 300$ K with the TDEP~\cite{Hellman2013} package in the relaxation time approximation on a 70 x 70 x 70 q-point grid, assuming natural isotope distribution. The mean free paths is given by $|\mathbf{v}\tau|$, where $\mathbf{v}$ is the group velocity and $\tau$ is the scattering time. The latter is computed based on third-order force constants and isotope disorder scattering, while the group velocity is computed from the phonon dispersions. Details can be found in Ref.~\cite{fugallo2013ab}.

\end{document}